\documentclass[12pt,a4paper]{article}
\usepackage[bitstream-charter]{mathdesign}
\usepackage{latexsym,amsmath, amsthm, geometry}
\usepackage{graphicx}
\usepackage{subcaption}
\usepackage{authblk}
\title{Weighted hazard ratio estimation for delayed and diminishing treatment effect}
\author[1]{Bharati Kumar}
\author[2]{Jonathan W. Bartlett}
\affil[1,2]{Department of Mathematical Sciences, University of Bath}

\begin{document}
\maketitle
\begin{abstract}
    Non proportional hazards (NPH) have been observed in confirmatory clinical trials with time to event outcomes. Under NPH, the hazard ratio does not stay constant over time and the log rank test is no longer the most powerful test. The weighted log rank test (WLRT) has been introduced to deal with the presence of non-proportionality. We focus our attention on the WLRT and the complementary Cox model based on time varying treatment effect proposed by Lin and León (2017) (doi: 10.1016/j.conctc.2017.09.004). We will investigate whether the proposed weighted hazard ratio (WHR) approach is unbiased in scenarios where the WLRT statistic is the most powerful test. In the diminishing treatment effect scenario where the WLRT statistic would be most optimal, the time varying treatment effect estimated by the Cox model estimates the treatment effect very close to the true one. However, when the true hazard ratio is large we note that the proposed model overestimates the treatment effect and the treatment profile over time. However, in the delayed treatment scenario, the estimated treatment effect profile over time is typically close to the true profile. For both scenarios, we have demonstrated analytically that the hazard ratio functions are approximately equal under certain constraints. In conclusion, our results demonstrates that in certain scenarios where a given WLRT would be most powerful, we observe that the WHR from the corresponding Cox model is estimating the treatment effect close to the true one.
\end{abstract}

\section{Introduction}
Randomised clinical trials (RCTs) are regarded as the gold standard for evaluating the effectiveness of a new treatment. In RCTs with time to event outcomes, the primary outcome is to measure the differences in the survival curves of the different arms and report the treatment effect to quantify treatment differences. The analysis of such trials are routinely accomplished by the log rank test and the Cox proportional hazard (PH) model. The hazard ratio is obtained from the Cox PH model to measure the effectiveness of the new treatment compared to the control group. The Cox PH model assumes that the hazard ratio between the treatment and control group stays constant from the start of study period until the end of the follow up period. Under PH, the log rank test is the most powerful test to detect differences in the survival curves \cite{FH}. However, non proportional hazards (NPH) are commonly observed in RCTs and have been reported in many published trials e.g. the IPASS trial in lung cancer \cite{mok2009gefitinib} and the ICON7 trial in ovarian cancer \cite{kristensen2011result} . When the PH assumption does not hold, the log rank test is still valid but may suffer substantial power loss and the interpretation of the hazard ratio becomes invalid and challenging \cite{hernan2010hazards}. There may be many reasons for non PH to exist in clinical trials, for instance due to delayed clinical effect in the case of immuno-oncology trials. 

Various statistical methods have been proposed to analyse time to event outcomes in clinical trials under non proportional hazards scenarios including restricted mean survival time (RMST) \cite{royston2013restricted}, weighted log rank tests \cite{fleming1981class} and maximum combination test (MaxCombo) \cite{lin2020alternative}. The RMST measures the average survival time from time $t= 0$ to some pre-specified time horizon $t=\tau$ and it may be estimated as the area under the survival curve up to that time point. The weighted log rank test based on the Fleming-Harrington class of weights $G^{\rho,\gamma}$ with weight $S(t)^{\rho}$ $(1-S(t))^{\gamma}$ where $S(t)$ is the survival function of the pooled patients from control and treatment group; the parameters $\rho$, $\gamma$ allows one to down weight early, late or middle events. The MaxCombo test uses the Fleming-Harrington weight function to analyse time to event outcomes in the presence of non PH. The test avoids having to pre-specify the weight function and considers a procedure which selects the best test from the set of test statistics. The procedure consist of selecting the best combination of weighted log rank test statistics with different choices of $\rho$ and $\gamma$.  When the MaxCombo approach is used, Lin et al \cite{lin2020alternative} suggest using the weighted hazard ratio for effect estimation. The estimated effect is obtained from a Cox model with a time-varying effect, where the weights are those associated with the weighted log rank test \cite{lin2017estimation}.

This paper focuses on the estimation of treatment effect under non proportional hazards, particularly the weighted hazard ratio method proposed by Lin and León \cite{lin2017estimation}. They propose fitting a particular Cox model with a time-varying covariate/effect for which the score test corresponds to a weighted log rank test. The model proposes an effect adjustment factor $A(t)$ = $\frac{w(t)}{max(w(t))}$ where $w(t)$ is the weight function from the chosen weighted log rank test. This is incorporated in to the Cox PH model to provide time varying effect which can be viewed as the treatment coefficient $\beta$ weighted by the adjustment factor $A(t)$. The hazard ratio obtained from the model is expressed as $HR(t) =e^{\beta A(t)}$ = $[HR^{F}]^{A(t)}$ where $HR^{F}$ represents the maximal effect at time $t$ with $A(t)=1$. In this paper, we are interested in investigating whether the estimated treatment effect is unbiased under scenarios in which the associated weighted log rank test is the most powerful. 

In Section 2, we describe the standard weighted log rank test and the proposed weighted hazard ratio estimation \cite{lin2017estimation}
and its estimation of the time-varying HR. In Section 3, we compare the hazard ratio functions obtained from the proposed model with the true hazard ratio functions analytically. In Section 4, we assess, through simulations, the performance of the proposed model under the two non PH scenario with different choices of $\rho$ and $\gamma$. Section 5 is a discussion.

\section{Weighted log rank tests and effect estimation}
In this section, we will provide a brief overview of the log rank test and the weighted log rank test, as well as introduce weighted hazard ratio for effect estimation proposed by Lin and León \cite{lin2017estimation}.

\subsection{Weighted log rank test}
Let $t_{1}$,$t_{2}$,\ldots,$t_{r}$ be the ordered failure times across both treatment arms, $d_{1j}$ and $d_{2j}$ be the number of deaths at $t_{j}$ in treatment and control arm respectively with $d_{j}$= $d_{1j}$ +$d_{2j}$  and $n_{1j}$ and $n_{2j}$ be the number of individuals at risk before $t_{j}$ with $n_{j}$ = $n_{1j}$ + $n_{2j}$. Let $X_{1}, X_{2},\dots,X_{r}$ denote the treatment assignment where $X_{j}$=1 if the j-th subject is assigned to treatment arm and $X_{j}$=0 if assigned to control arm. The log-rank test statistic is defined as, 
\begin{align*}
    \begin{split}
        Z= \frac{\sum_{j=1}^{r}(d_{1j} - e_{1j})}{(\sum_{j=1}^{r}V_{L})^{1/2}}
    \end{split},
\end{align*}
where $e_{1j}$= $\frac{n_{1j} d_{j}}{n_{j}}$ is the expected number of failure at time $t_{(j)}$ under the null hypothesis of equal survival curves. $V_{L}$ is the variance of the observed number of failures, $V_{L}$ = $ \sum_{j=1}^{r}\frac{n_{1j}n_{2j}d_{j}(n_{j}-d_{j})}{n_{j}^{2}(n_{j}-1)}$. Under the null hypothesis of equal survival functions in the two treatment groups, asymptotically the test statistic has a chi-squared distribution on one degree of freedom. Although the log rank test is still a valid test under non proportional hazards it is not the most powerful test \cite{royston2013restricted}. Therefore, the weighted version of log rank test is sometimes utilised when the ratio of the two hazard functions are not constant over time \cite{fleming1981class}. 

The weighted log rank test (WLRT) assigns a weight function to different time points depending on the expected type of non-proportional hazard scenario. For instance, in immuno-oncology trials there may be a delayed treatment effect and therefore the survival curve for the treatment group will only emerge to separate from the control survival curve after a certain period of time. In this case, higher weights can be allocated to later time points \cite{lin2020alternative}. Formally, the WLRT is defined as,

\begin{equation}
    \begin{split}
    \label{WLRTST}
        W_{K}=
        \frac{\sum_{j=1}^{r}w(t_{j})(d_{1j} - e_{1j})}{(\sum_{j=1}^{r}V_{L})^{1/2}} 
    \end{split},
\end{equation}
with variance
\begin{align*}
    \begin{split}
        V_{K} = \sum_{j=1}^{r}w(t(j))^{2}\bigg(\frac{n_{1j}n_{2j}d_{j}(n_{j}-d_{j})}{n_{j}^{2}(n_{j}-1)}\bigg)
    \end{split}
\end{align*}
where $w(\cdot)$ is the non-negative weight function. We will consider the Fleming-Harrington family of weighted log rank test, commonly denoted as $G^{\rho,\gamma}$ with weight,
\begin{equation}
\label{weight}
    w(t) = \hat{S}(t)^{\rho} (1-\hat{S}(t))^{\gamma}
\end{equation}
where $\hat{S}(t)$ is the Kaplan- Meier estimate of the survival function of the pooled patients from the treatment and control arm. The parameters $\rho$ and $\gamma$ control the shape of the weight function. $G^{0,0}$ represents the log-rank test with $w(t)=1$ that has constant weight over time; delayed treatment effects can be tested using $G^{0,\gamma}$ with $w(t)=(1-\hat{S}(t))^{\gamma}$ that allocates higher weight at later time points to detect late survival difference, $G^{1,0}$ represents diminishing effect with $w(t)=\hat{S}(t)$ that gives more weight the earlier time points to detect early separation; $G^{1,1}$ represents mid separation with $w(t)=\hat{S}(t) (1-\hat{S}(t))$ that puts more weight at the middle of the follow-up period than at the ends.

\subsection{Weighted hazard ratios}
Lin and León \cite{lin2017estimation} proposed fitting a Cox model to provide a time varying treatment effect estimate to complement the weighted log rank test. Lin and León propose fitting a Cox model with,
$$
\lambda(t;X) = \lambda_{0}(t)e^{A(t)\beta X}
$$
where $A(t)$ is defined by,
$$
A(t) = \frac{w(t)}{\textrm{max}(w(s))}
$$
 $\textrm{max}(w(s))$ is evaluated over the values of $t$ in the observed dataset and $A(t)$ thus has a maximum value of 1. The time varying covariate $X^{*}(t) = A(t)X$ represents the treatment assignment weighted by the adjustment factor. Once $X^{*}(t)$ has been determined, the constant coefficient $\beta$ can be estimated.
It is shown in the paper that the score test of the null hypothesis that $\beta=0$ in the proposed model above is equivalent to the weighted log rank test with the corresponding choice of weight function \cite{lin2017estimation}. Lin and León used the fact that the score test from this Cox model corresponds to a particular weighted log rank to motivate estimating the time-varying treatment effect using this Cox model. The coefficient $\beta$ can be interpreted as the maximal effect where the model assumes the patients experience the most benefit ($A(t)=1$). To differentiate the time varying hazard ratio derived from the proposed method, we will denote this as $HR_{LL}$ and it can be expressed as,
\begin{equation}
 HR_{LL}(t)= \frac{\lambda_{0}(t)e^{A(t)\beta\times1}}{\lambda_{0}(t)e^{A(t)\beta\times0}} = e^{\beta A(t)} = \big[HR^{F}\big]^{A(t)}
 \label{HRlin}
\end{equation}
where $HR^{F}$=$e^{\beta}$ represents the maximal effect (i.e., at time $t$ with $A(t)$=1). The time profile of the treatment effect can be observed as the treatment coefficient $\beta$ weighted by the adjustment factor $A(t)$. The estimate derived from the model provides a consistent time profile of the hazard ratio over time given that the true hazard ratio also varies over time as specified by $A(t)$. If this does not hold then the treatment effect measure will in general no longer be unbiased for the true value of the hazard ratio at a given time.

\section{Evaluation of weighted hazard ratio estimation method}

In this section we give expressions for the hazard ratio function for which the Fleming-Harrington \cite{FH} $G^{\rho,\gamma}$ test statistics are most efficient.
 We will also be exploring whether, for a given weight function, the Lin and León estimation method \cite{lin2017estimation} gives you the correct treatment effect profile if the true data generating mechanism is such that the corresponding weighted log rank test is the most powerful.
\subsection{Diminishing effect}

Fleming and Harrington \cite{FH} showed that the $G^{\rho}$ test statistic is the most powerful test statistic when the hazard ratio at time $t$ is of the following form,
\begin{equation}
\label{eqn: efficient}
\frac{\lambda_{2}(t)}{\lambda_{1}(t)}=\frac{e^{\Delta}}{\{S_{1}(t)\}^{\rho} + e^{\Delta}[1-\{S_{1}(t)\}^{\rho}]}
\end{equation}
where $S_1(t)=\exp\left(-\int_{0}^{t}\lambda_{1}(x) \textrm{dx}\right)$ is the survival function for the control arm. The parameter $e^{\Delta}$ represents the hazard ratio at $t=0$. The parameter $\rho$ controls how quickly the effect diminishes. The above hazard ratio function \eqref{eqn: efficient} has its full treatment effect initially $e^{\beta}$ and decreases monotonically to 1. Later, we will demonstrate various examples with $\rho =0.5,1,2$ to further develop our understanding about the behaviour of the weighted logrank test and the proposed model.
We now calculate the weighted hazard ratio function proposed by Lin and León \cite{lin2017estimation} under the set up where the $G^{\rho}$ test statistic would be optimal. In order to calculate the proposed weighted hazard ratio function under the set up where WLRT is optimal, we will need to derive the true pooled survival function which is determined by finding the survival function of the two treatment arms. The hazard function for the treatment arm $\lambda_{2}(t)$ can be derived, assuming that the true data generating mechanism is such that that WLRT is optimal, by using equation \eqref{eqn: efficient}
\begin{align*}
\begin{split}
\lambda_{2}(t)& =\frac{\lambda_{1}(t)e^{\Delta}}{\{S_{1}(t)\}^{\rho} + e^{\Delta}[1-\{S_{1}(t)\}^{\rho}]} \\
    \end{split}
\end{align*}
Thus, the survival function for the treatment arm is,
\begin{equation}
    \begin{split}
    \label{survivalTRT}
        S_{2}(t) &= \exp\left(-\int_{0}^{t}\lambda_{2}(x) \textrm{dx}\right) \\
        &= \exp\left(-\int_{0}^{t} \frac{\lambda_{1}(x)e^{\Delta}}{\{S_{1}(x)\}^{\rho} + e^{\Delta}[1-\{S_{1}(x)\}^{\rho}]} dx\right)
    \end{split}
\end{equation}
We can replace the adjustment factor $A(t)$ with the weight function that the Lin and León method converges to (as $n$ tends to $\infty$) \eqref{HRlin}, in this case $S(t)$. For the Lin and León \cite{lin2017estimation} method, we estimate the pooled survival function $S(t)^{\rho}$ by the Kaplan-Meier estimator for the pooled sample which is consistent for the true pooled survival function. Therefore the Lin and León method consistently estimates,
\begin{equation}
\label{eqn:hr}
    \begin{split}
    HR_{LL}(t) = e^{\beta A(t)} = e^{\beta\big(0.5S_{1}(t) + 0.5S_{2}(t)\big)^{\rho}}
    \end{split}
\end{equation}
Substituting $S_{2}(t)$ in the hazard ratio function \eqref{eqn:hr} for which this WLRT is optimal for gives,
\begin{equation}
\begin{split}
\label{LLeff}
    HR_{LL}(t) = \textrm{exp}\left[\beta \left(\frac{1}{2}S_{1}(t)\right) +\frac{1}{2}\exp\left(-\int_{0}^{t} \frac{\lambda_{1}(x)e^{\Delta}}{\{S_{1}(x)\}^{\rho} + e^{\Delta}[1-\{S_{1}(x)\}^{\rho}]} dx\right)\Bigg)^{\rho}\right]
\end{split}
\end{equation}
When fitting the proposed model, the $\beta$ is estimated from the Cox PH model once the adjustment factor $A(t)$ is determined. 
Comparing the true hazard ratio function \eqref{eqn: efficient} with the proposed hazard ratio function \eqref{LLeff}, we see that the two expressions do not appear to be equal to each other. In Appendix B, we demonstrate that the two hazard ratio functions are approximately equal to each other when $\Delta$ is close to 0. We also demonstrate this equality graphically in the next section where we choose the value of $e^{\Delta}$ to compare the hazard ratio functions.
We also compare the hazard ratio expressions analytically for some particular choices of $\lambda_1(t)$ and $e^{\Delta}$ to illustrate in Appendix A.

\subsection{Delayed treatment effect}
 The paper by Garès \cite{gares2017fleming} has provided an expression for the hazard function for which the Fleming-Harrington test $G^{\gamma}$ for detecting late effects is optimal. Assume, the survival time in the control arm is exponentially distributed with some constant hazard $\lambda_{1}$. Then Garès et al \cite{gares2017fleming} show that the $G^{\gamma}$ test is optimal for testing the hypothesis,

\begin{equation}
\begin{split}
H_{0}: \lambda_{CT}&= \lambda_{TRT}= \lambda_{1}\\
   H_{1}: \lambda_{TRT}(t)&=\lambda_{CT} \frac{L^{\gamma}((\mathcal{L^{\gamma}})^{-1}(\mathcal{L^{\gamma}}(e^{-\lambda_{1}t})+\varphi))}{L^{\gamma}(e^{-\lambda_{1}t})}
   \label{haz2}
   \end{split}
\end{equation}
where $L(x)=\int_{x}^{1}\frac{(1-s)}{s} ds= 1
+\ln(x)-x$ and $\mathcal{L}(x)=\int_{0.5}^{x} \frac{1}{s\log(s)+s-s^{2}} ds$. The expression for the hazard ratio has no closed form and needs numerical integration techniques to compute. Garès et al \cite{gares2017fleming} has provided the definition for the parameter $\varphi$ which is evaluated at some chosen time point $\tau$ and is of the following form,
\begin{equation}
\begin{split}
\label{delta}
   \varphi&= \mathcal{L}^{\gamma}(r(1-S_{1}(\tau))+S_{1}(\tau))-\mathcal{L}^{\gamma}(S_{1}(\tau)) \\
   &= \mathcal{L}^{\gamma}(S_{2}(\tau))- \mathcal{L}^{\gamma}(S_{1}(\tau))
   \end{split}
\end{equation}
where $r$ is the discrepancy rate, $r=\frac{S_{2}(\tau)-S_{1}(\tau)}{1-S_{1}(\tau)}$. 

Similar to the diminishing effect scenario, we can check whether the true hazard ratio for which the $G^{\gamma}$ test is optimal, as given in equation \eqref{haz2}, agrees with the hazard ratio estimation proposed by Lin and León \cite{lin2017estimation}. 
The true hazard ratio under which the $G^{\gamma}$ test statistic is optimal is shown to be the second equation in \eqref{haz2}. Now the the weighted hazard ratio model proposed by Lin and León corresponds to, 
\begin{align*}
    \begin{split}
 HR_{LL}(t) =e^{\beta A(t)} &= e^{\beta \frac{w(t)}{max(w(s))}}\\
       &= e^{\beta \left(\frac{\left(1-S(t)\right)^{\gamma}}{max(w(s))}\right)}=e^{\beta\left(\frac{(1-0.5S_{1}(t)-0.5S_{2}(t))^{\gamma}}{max(w(s))}\right)}
    \end{split}
\end{align*}
From Garès et al, the survival in the treatment and control arm is defined as $S_{2}(t)=(\mathcal{L})^{-1}(\mathcal{L}(e^{-\lambda_{1}t})    +\mathcal{L}(S_{2}(\tau))-\mathcal{L}(S_{1}(\tau)))= (\mathcal{L})^{-1}(\mathcal{L}(e^{-\lambda_{1}t})+ \varphi)$ and $S_{1}(t)=e^{-\lambda_{1}}t$ \cite{gares2017fleming}. We now substitute in for $S_{1}(t)$ and $S_{2}(t)$,

\begin{equation}
    \begin{split}
 HR_{LL}(t)&=e^{\beta \left( \frac{\left(1-0.5e^{-\lambda_{1}t}-0.5\mathcal{L}^{-1}(\mathcal{L}(e^{-\lambda_{1}t})+\varphi)\right)^{\gamma}}{\left(1-0.5e^{-\lambda_{1}\tau}-0.5\mathcal{L}^{-1}(\mathcal{L}(e^{-\lambda_{1}\tau})+\varphi)\right)^{\gamma}}\right)}
        \label{LLHR}
    \end{split}
  \end{equation}
We assume that the Lin and León method estimates the correct hazard ratio ($e^{\beta}$) at the time point in the study where it is the biggest effect i.e. at $t=\tau$. Like in Section 3.1, we show that if the Lin and León method estimates the correct hazard ratio at the time point when it is largest, then it will estimate the HR at all other times correctly too. In Appendix C, we see that the expression for the hazard ratio function from Lin and León model \eqref{LLHR} does approximately match the true HR assumed as per equation \eqref{haz2} under the condition that $\varphi$ is small. 

\section{Simulation studies}
This section investigates the weighted hazard ratio method under diminishing and delayed effect using simulations \cite{lin2017estimation}. We will focus on investigating the method proposed by Lin and León and its properties when non proportional hazards are present in the data. Specifically, we will empirically check the analytical results of Section 3 that showed that the Lin and León approach is approximately unbiased when the treatment effect is small, and how large the bias might be when the treatment effect is larger.

\subsection{Scenario I: Diminishing effect}
We simulated the survival data where the treatment is assumed to have its maximal effect initially and decrease monotonically to 1. The survival function for the treatment arm is given in equation \eqref{survivalTRT}.  Three different scenarios $\rho=0.5,1,2$ were considered to understand the behaviour of the proposed model under various rates at which the effect diminishes. We also chose three different values of the maximum effect $e^{\Delta}=1.4,4,8$ to investigate whether substantial bias occurs when the treatment effect is larger. We fitted the Cox model proposed by \cite{lin2017estimation} where the weight function $w(t)=\hat{S}(t)^{\rho}$ for the non PH data we simulated is specified for three scenarios $\rho=0.5,1,2$. The study enrolls 200 patients, equally allocated in the two arms. The survival of the control arm follows an exponential distribution with hazard rate $\lambda_{1}=0.5$. The survival in the treatment arm is as per equation \eqref{survivalTRT} . The follow up time is assumed to be 3 years; patients whose event time exceeds 3 years are censored. We conduced 5000 runs for each scenario.

Figure \ref{fig:hazard} shows the true hazard ratio over time for which the $G^{\rho}$ statistic is optimal and the average hazard ratio profile over time estimated by the Lin and Leòn method. The plot displays $e^{\bar{\hat{\beta}}S(t)^{\rho}}$ where $\bar{\hat{\beta}}$ is the average of the hazard ratio at $t=0$ estimated from the Lin and Leòn method and $S(t)$ is the true pooled survival function. The proposed model overestimates the maximal effect when the true $HR$ at $t=0$ i.e. $e^{\Delta}$ is large, and slightly underestimates immediately after 1 month of follow up. Note that when the treatment effect is small, the proposed model estimates the maximal effect very close to the true one, in line with our analytical results in Appendix B. We observe through our work that when the treatment effect is large, the proposed model is not able to estimate the maximal effect at $t=0$ correctly which has an impact on the treatment effect profile over time, in that case the implied profile of treatment effect over time would not match the true one (as given by equation \eqref{eqn: efficient}). 
\vspace{3cm}

\begin{figure}[h!]
     \centering
     \begin{subfigure}[b]{0.7\textwidth}
         \centering
     \includegraphics[width=0.9\textwidth]{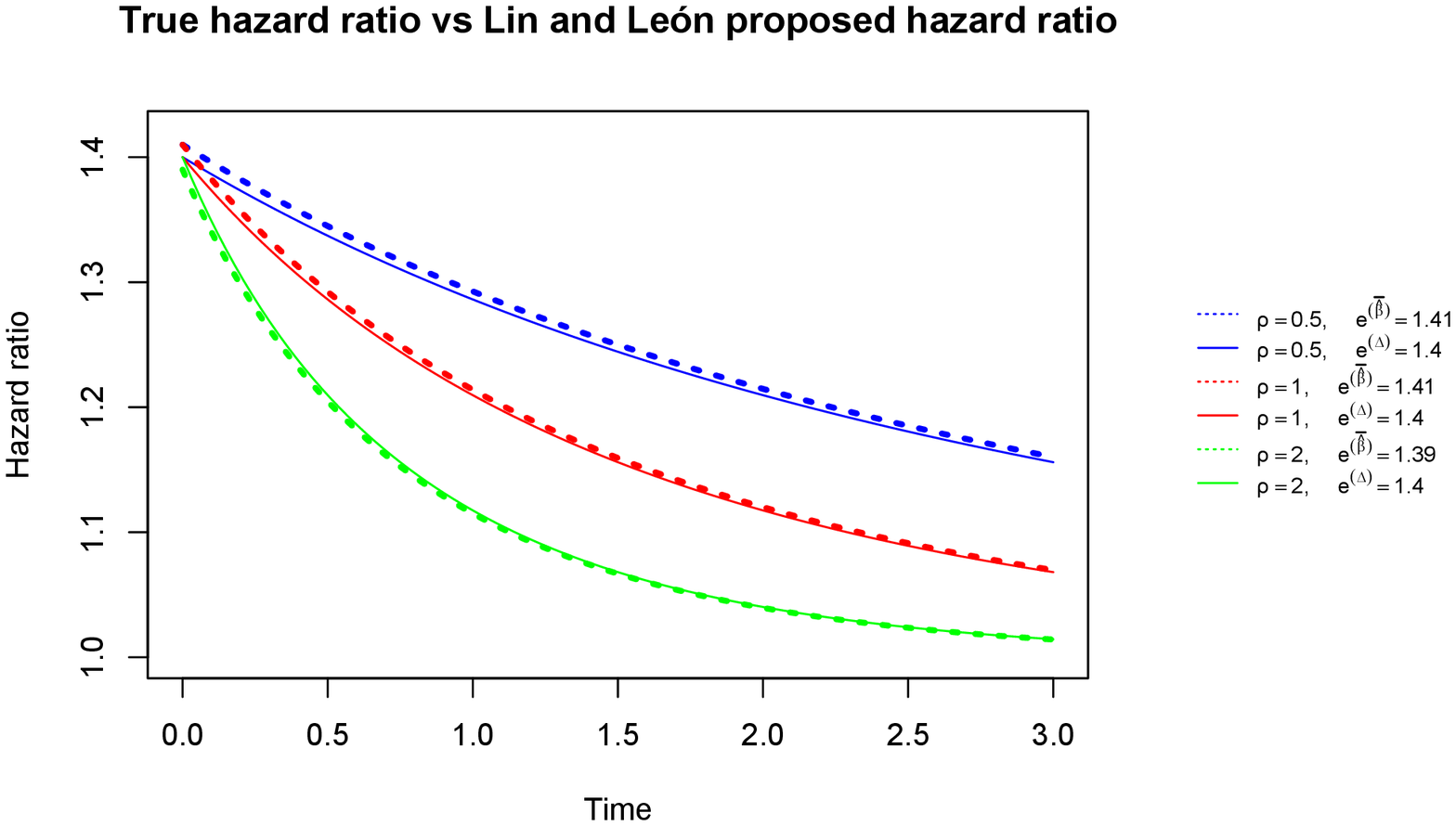}
         \caption{$e^{\Delta}=1.4$}
         \label{fig:y equals x}
     \end{subfigure}
     \hfill
     \begin{subfigure}[b]{0.7\textwidth}
         \centering
     \includegraphics[width=0.9\textwidth]{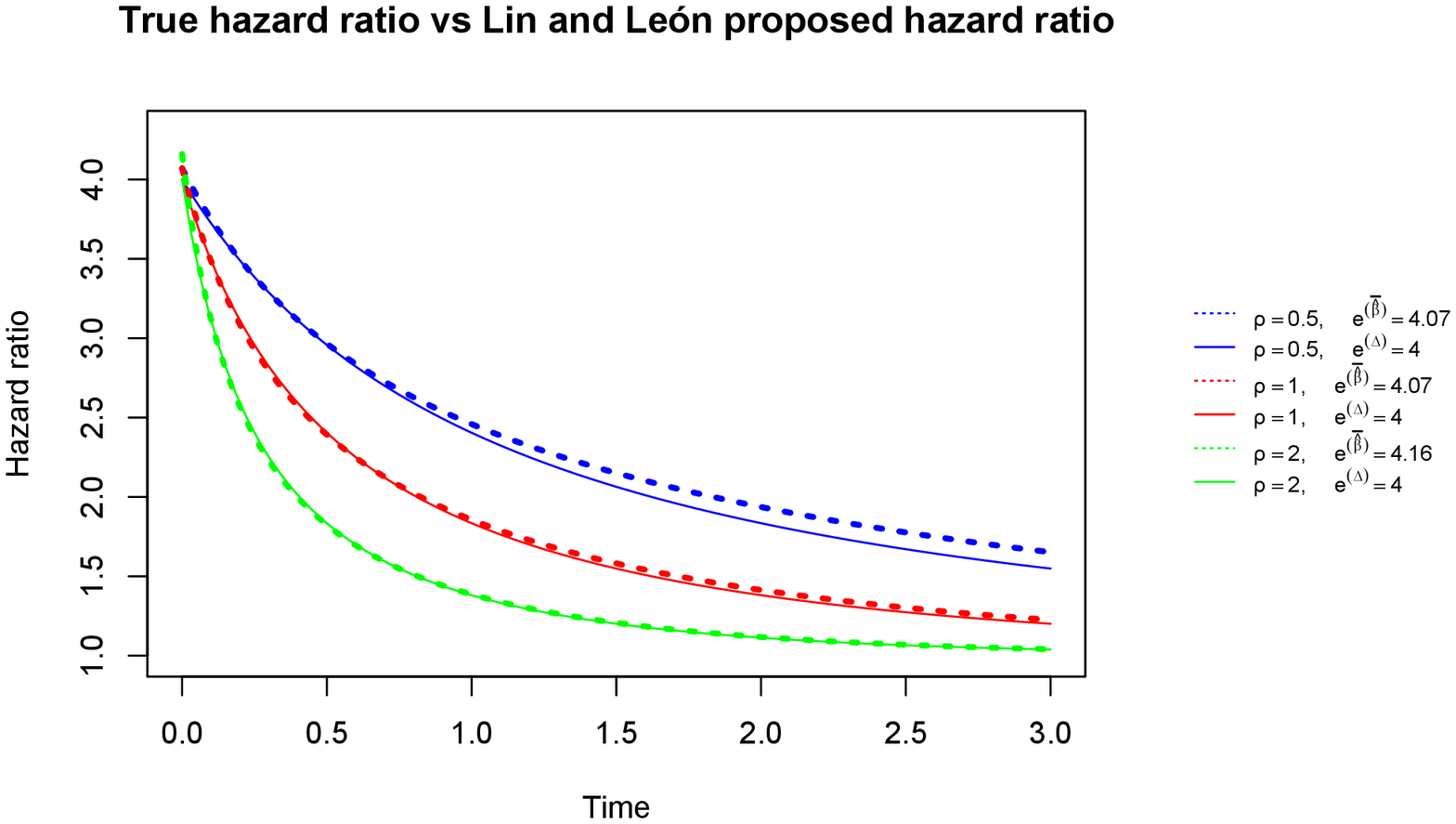}
         \caption{$e^{\Delta}=4$}
         \label{fig:three sin x}
     \end{subfigure}
     \end{figure}
     \begin{figure}[ht]\ContinuedFloat
     \centering
     \begin{subfigure}[b]{0.7\textwidth}
         \centering
     \includegraphics[width=0.9\textwidth]{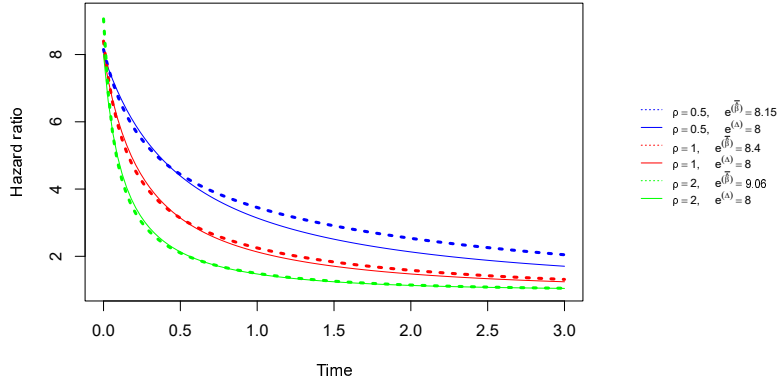}
         \caption{$e^{\Delta}=8$}
         \label{fig:five over x}
     \end{subfigure}
        \caption{Hazard ratio over time targeted by weighted hazard ratio model superimposed true hazard ratio over time under a data generating mechanism where the $G^{\rho}$ test is optimal. The solid line shows the true hazard ratio over time for $\rho=0.5,1,2$. The true hazard ratio for the three scenarios $\rho=0.5,1,2$ at $t=0$ are $e^{\Delta}=1.4,4,8$ respectively and the hazard rate in the control arm stays the same in all the scenarios $\lambda_{1}(t)=0.5$. The dotted lines show the proposed hazard ratio targeted by the weighted hazard ratio model. The hazard ratio for the dotted lines is the average of the hazard ratios estimated from fitting the Lin and León method and is as follows; the average of the log hazard ratio across 5000 simulations is $e^{\bar{\hat{\beta}}}=1.41,4.07,8.15$,$e^{\bar{\hat{\beta}}}=1.41,4.07,8.40$, $e^{\bar{\hat{\beta}}}=1.39,4.16,9.06$ for $\rho=0.5,1,2$.}
        \label{fig:hazard}
\end{figure}

\newpage
\subsection{Scenario II: Delayed treatment effect}
In this scenario, the data were simulated based on the assumption that the treatment will have no effect at the start of the follow up and gradually increases to have its maximal effect at the final time point of follow up period. The study enrolls the same number of patients as in the diminishing effect scenario i.e. 200. The maximal treatment effect occurs at the final time point $\tau=2$. 
The survival times in the control arm are exponentially distributed with hazard rate 0.5.  The survival times in the active arm can then be simulated using the expression for the hazard function given in \eqref{haz2}. To simulate the survival times in the active arm, we calculated the value of $S_{2}(t)$ for $t$ from 0 to $\tau$ in increments of 0.0005. We then simulated $U \sim U(0,1)$ and found the value of $t$ such that $S_{2}(t)$ was closest to $U$. We consider two scenarios of  $S_{2}(\tau)=0.25$ and 0.1 which provides the values of discrepancy rate $r=-0.19$ and $-0.42$ for $\gamma=0.5,1,2$ to verify our results obtained in Section 3. The mechanism behind simulating the non proportional data in this way would ensure that the weight function $(1-S(t))^{\gamma}$ would provide the most powerful test statistic. Similar to the previous scenario, we applied the Lin and León Cox PH model to our simulated datasets and compared the estimated HR profile with the true HR profile over time. We again considered three scenarios of $\gamma$ where it is varied to control the rate of how quickly the treatment effect reaches its maximum effect. 

\begin{figure}[hb!]
     \centering
     \begin{subfigure}[b]{0.7\textwidth}
         \centering
     \includegraphics[width=0.9\textwidth]{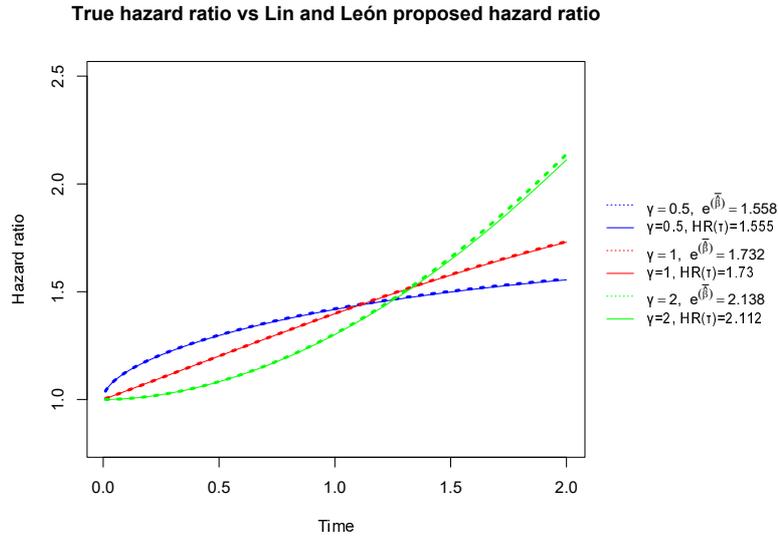}
         \caption{$S_{2}(\tau)=0.25$}
         \label{fig:sc2}
     \end{subfigure}
     \hfill
     \begin{subfigure}[b]{0.7\textwidth}
         \centering
     \includegraphics[width=0.9\textwidth]{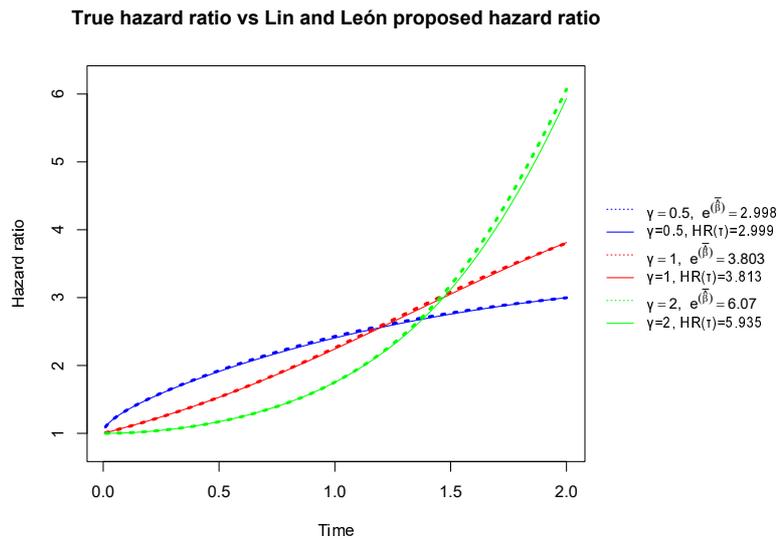}
         \caption{$S_{2}(\tau)=0.1$}
         \label{fig:sc3}
     \end{subfigure}
    \caption{Hazard ratio over time targeted by weighted hazard ratio model superimposed true hazard ratio over time under a data generating mechanism where the $G^{\gamma}$ test is optimal. The solid line shows the true hazard ratio over time for $\gamma=0.5,1,2$. The true HR for $r=-0.19$ at $\tau=2$ is $e^{\beta}=1.555,1.73,2.112$ and for $r=-0.42$ is $e^{\beta}=2.999,3.813,5.935$ for $\gamma=0.5,1,2$ respectively. The dashed lines show the proposed hazard ratio targeted by the weighted hazard ratio model. The hazard ratio for the dashed lines is the average of the hazard ratios across 5000 simulations estimated from fitting the Lin and León method and is as follows; $e^{\bar{\hat{\beta}}}=1.558,1.732,2.138$, $e^{\bar{\hat{\beta}}}=2.998,3.803,6.070$ for $r=-0.19,-0.42$ and $\gamma=0.5,1,2$ respectively}
    \label{fig:hazard_ratioplot}
\end{figure}

The true hazard ratio over time for which the $G^{\gamma}$ statistic is optimal and the average hazard ratio profile over time estimated by the Lin and Leòn method is displayed in Figure \ref{fig:hazard_ratioplot}. The plot displays the three HR profiles over time for $\gamma=0.5,1,2$ for two different values of $S_{2}(\tau)$.
 The solid lines represents the true HR profile and the dashed lines represents $e^{\bar{\hat{\beta}}(1-S(t)^{\gamma}}$. Note that for both values of $S_{2}(\tau)$, the estimated average hazard ratio is overestimated for $\gamma=1,2$ and this deviation is reflected on the HR time profile. Though for other scenarios, we see that the Lin and León method estimates the treatment effect profile over time with minimal bias, in line with our analytical results in Appendix C.The proposed model closely estimates the maximal treatment effect for both scenarios where $\gamma=0.5$ and $r$ is relatively small.

\newpage
\section{Discussion}
 In this paper, we considered the Fleming-Harrington class of weights to simulate non-proportional hazard data for the which the same weight function would give the most optimal test. The Lin and León method consists of fitting a Cox model based on a time-varying treatment covariate. The score test from the proposed model gives a $p$-value which is equivalent to the weighted log rank test and the estimate
derived from the model provides a time-profile of the treatment effect. We considered two non proportional hazard scenarios to investigate its ability to correctly estimate the maximal treatment effect and the corresponding treatment effect profile over time when the data were simulated such that the corresponding weighted log rank test would be most powerful.

We showed analytically that the Lin and Leòn method is approximately unbiased in both scenarios when the treatment effect is small. Our simulation results for the diminishing treatment effect scenario showed that the Lin and León method overestimates the maximal treatment effect when the true treatment effect at $t=0$ is large as well as when $\rho$ is large. For scenarios where the treatment effect is small, the proposed model estimates the maximal effect very close to the true one. For the delayed treatment effect scenario, the Lin and León method overestimates the maximal effect for $\gamma=1,2$ but in our simulations not by very much. The Lin and León method gave essentially unbiased estimates of the full effect and treatment effect profile over time for $\gamma=0.5$.

We have shown through simulations that under the delayed treatment effect scenarios, the proposed model is giving close to the correct profile over time when treatment effect is small. However, this does not guarantee that the proposed model would provide unbiased results had the data generating mechanism been different, for example if the the hazard in the control arm was not constant and changes over time. 

The choice of $\rho$ and $\gamma$ in estimating the treatment effect requires extensive knowledge of how the true hazard ratio changes over time. From a testing perspective, choosing the correct value of $\rho$ or $\gamma$ is less crucial since the test still controls the type I error rate if the incorrect value is used. However, for estimation of the treatment effect profile over time, use of the incorrect value would lead to biased estimates. Given the difficulty of knowing the correct value of $\rho$ and $\gamma$, it may be preferable to use methods such as restricted mean survival time which do not rely on how quickly or slowly the treatment effect changes over time, and for which the interpretation of the treatment effect is clinically meaningful to the clinicians and patients \cite{royston2013restricted}. 
\section*{Acknowledgement}
Bharati Kumar’s research was funded by a UK EPSRC studentship (2281147). Jonathan Bartlett was supported by the UK MRC (grant MR/T023953/1).
\appendix

\section*{\\Appendix A: Computation of the true  hazard ratio function and proposed hazard ratio function for the diminishing effect scenario for general value of $\rho$}

We let $\lambda_{1}(t)= \lambda_{1}$ and calculate the true hazard ratio function \eqref{eqn: efficient} and the proposed hazard ratio \eqref{LLeff}. The survival function for the control arm,
\begin{align*}
    S_{1}(t) = \textrm{exp}\left(-\int_{0}^{t}\lambda_{1}(x)\right) \textrm{dx} = \textrm{exp}(-\lambda_{1}t)
\end{align*}
The hazard function for the treatment arm $\lambda_{2}(t)$ can be derived by the relationship in equation (\ref{eqn: efficient})
\begin{align*}
    \begin{split}
        \lambda_{2}(t)&= \lambda_{1} \times e^{\Delta}\left[\exp(-\lambda_{1}t)^{\rho} + [1-\{\exp(-\lambda_{1}t)\}^{\rho}]\times e^{\Delta}\right]^{-1}\\
        &= \lambda_{1} \times e^{\Delta}\left[e^{\Delta}+\exp(-\lambda_{1}t)^{\rho} \left\{1-e^{\Delta}\right\}\right]^{-1}
    \end{split}
\end{align*}
Thus, the survival function for the treatment arm is,
\begin{align*}
    \begin{split}
        S_{2}(t) &= \exp\left(-\int_{0}^{t}\lambda_{2}(x) \textrm{dx}\right) \\
      \int_{0}^{t}\lambda_{2}(x)\textrm{dx} &= \int_{0}^{t}\lambda_{1} \times e^{\Delta}\left[e^{\Delta}+\exp(-\lambda_{1}x)^{\rho} \left\{1-e^{\Delta}\right\}\right]^{-1} \textrm{dx}\\
      &=\left[\frac{1}{\rho}\log\left(1-e^{\Delta}+e^{\Delta} \{\exp(-\lambda_{1}x)\}^{\rho}\right)\right]_{0}^{t}\\
      S_{2}(t)&=\exp\left(-\frac{1}{\rho}\log\left(1-e^{\Delta}+e^{\Delta} \{\exp(-\lambda_{1}t)\}^{\rho}\right)\right)
      \\
     &=\frac{1}{\left[1-e^{\Delta}+e^{\Delta}\{\exp(\lambda_{1}t)\}^{\rho}\right]^{1/\rho}}
    \end{split}
\end{align*}

Recall, the model proposed by Lin and Leòn \cite{lin2017estimation} states that the hazard ratio between the two arms can be expressed as, 
\begin{align*}
    \begin{split}
        HR_{LL}(t) = \frac{\lambda_{0}e^{A(t)\beta\times1}}{\lambda_{0}e^{A(t)\beta\times0}} = e^{\beta A(t)}
    \end{split}
\end{align*}
Let us substitute the pooled survival function, instead of $A(t)$,
\begin{equation}
    \begin{split}
    \label{apphr}
    HR_{LL}(t) = e^{\beta A(t)} = e^{\beta\left(0.5S_{1}(t) + 0.5S_{2}(t)\right)^{\rho}}
    \end{split}
\end{equation}
For our example, the true hazard ratio under our data generating mechanism is expressed as,
\begin{align*}
    \begin{split}
        HR(t) &=\frac{e^{\Delta}}{e^{\Delta}+e^{\Delta}\left(1-\{\exp(-\lambda_{1} t)\}^{\rho}\right)}
    \end{split}
\end{align*}
Now the hazard ratio described in \eqref{apphr} is expressed as,
\begin{align*}
    \begin{split}
        \frac{1}{2}S_{1}(t) + \frac{1}{2}S_{2}(t)& = \frac{1}{2}\exp(-\lambda_{1}t) + \frac{1}{2}\frac{1}{\left[1-e^{\Delta}+e^{\Delta}\{\exp(\lambda_{1}t)\}^{\rho}\right]^{1/\rho}} \\
          HR_{LL}(t)&= \exp\left[\beta\cdot \left(\frac{1}{2}\exp(-\lambda_{1}t) + \frac{1}{2}\frac{1}{\left[1-e^{\Delta}+e^{\Delta}\{\exp(\lambda_{1}t)\}^{\rho}\right]^{1/\rho}}\right)^{\rho}\right] 
    \end{split}
\end{align*}

\section*{\\Appendix B: Taylor series expansion of the true hazard ratio function and the proposed hazard ratio function for the diminishing effect}

In this section, we demonstrate that the true hazard ratio function and the proposed hazard ratio functions are approximately equal when $e^{\Delta}$ is close to 1 i.e., when $\Delta$ is small. The evaluation of these results are executed generally for $\rho$. Recall the true hazard ratio function and the proposed hazard ratio function are of the following form,

\begin{align*}
    \begin{split}
        HR(t,\Delta) &=\frac{e^{\Delta}}{S_{1}(t)^{\rho}+e^{\Delta}\left(1-S_{1}(t)^{\rho}\right)} \\
        HR_{LL}(t,\Delta)&= \exp\left[\beta\cdot \left(\frac{1}{2}S_{1}(t) + \frac{1}{2}S_{2}(t)\right)^{\rho}\right] 
    \end{split}
\end{align*}

We first assume $\beta=\Delta$ in order to check whether the Lin and León model estimates the treatment effect over time correctly if it estimates the initial effect correctly. 

Suppose we let $f(\Delta)=\frac{HR(t,\Delta)}{HR_{LL}(t,\Delta)}$. We now proceed with a Taylor series expansion for $f(\Delta)$, 
\begin{align*}
    f(x)=\sum_{n=0}^{\infty}\frac{f^{(n)}(a)}{n!}(x-a)^{n}=f(a)+f'(a)(x-a)+f''(a)\frac{(x-a)^{2}}{2!}+\cdot\cdot\cdot
\end{align*}
We shall expand the first two terms of the series about $\Delta=0$, 
\begin{align*}
    f(\Delta)\approx f(0)+f'(0)(\Delta-0)
\end{align*}
When $a=0$, $f(0)$=$\frac{HR(t,0)}{HR_{LL}(t,0)}$=1. We now find the first order derivative of $f(\Delta)$ with respect to $\Delta$ using the quotient rule, so we have
\begin{align*}
   \frac{HR_{LL}(t,0)\frac{\partial HR(t,0)}{\partial \Delta}-HR(t,0)\frac{\partial HR_{LL}(t,0)}{\partial \Delta}}{(HR_{LL}(t,0))^{2}}=\frac{\partial HR(t,0)}{\partial \Delta}-\frac{\partial HR_{LL}(t,0)}{\partial \Delta}
\end{align*}

We derive the following derivatives for the true hazard ratio function using the product and chain rule,
\begin{align*}
\begin{split}
    \frac{\partial HR(t,\Delta)}{\partial \Delta}&=\frac{e^{\Delta}}{S_{1}(t)^{\rho}+e^{\Delta}\left(1-S_{1}(t)^{\rho}\right)}-\frac{e^{\Delta}\left(e^{\Delta}\left(1-S_{1}(t)^{\rho}\right)\right)}{\left(S_{1}(t)^{\rho}+e^{\Delta}\left(1-S_{1}(t)^{\rho}\right)\right)^{2}}
    \\
   \frac{\partial HR(t,0)}{\partial \Delta}&=\frac{1}{S_{1}(t)^{\rho}+1-S_{1}(t)^{\rho}}-\frac{1-S_{1}(t)^{\rho}}{\left(S_{1}(t)^{\rho}+1-S_{1}(t)^{\rho}\right)^{2}}=S_{1}(t)^{\rho}
    \end{split}
\end{align*}
Now, we find the derivative for the proposed hazard ratio function,
\begin{align*}
    \begin{split}
       HR_{LL}(t,\Delta)&= \exp\left[\Delta\cdot \left(\frac{1}{2}S_{1}(t) + \frac{1}{2}S_{2}(t,\Delta)\right)^{\rho}\right], 
    \end{split}
\end{align*}
where we now emphasise the dependence of $S_{2}(t,\Delta)$ on $\Delta$.
\begin{align*}
\begin{split}
         \frac{\partial HR_{LL}(t,0)}{\partial \Delta} &=\left\{\left(\frac{1}{2}S_{1}(t) + \frac{1}{2}S_{2}(t,\Delta)\right)^{\rho} +\Delta\rho\left[ \left(\frac{1}{2}S_{1}(t) + \frac{1}{2}S_{2}(t,\Delta)\right)^{\rho-1} \left(\frac{1}{2}\frac{\partial S_{2}(t,\Delta)}{\partial \Delta} \right)\right] \right\} \times \\
 &\exp\left[\Delta\cdot \left(\frac{1}{2}S_{1}(t) + \frac{1}{2}S_{2}(t,\Delta)\right)^{\rho}\right]
    \end{split}
\end{align*}
We now evaluate the derivative of the proposed hazard ratio function at $\Delta=0$,
\begin{align*}
    \begin{split}
         \frac{\partial HR_{LL}(t,0)}{\partial \Delta} &=\left(\frac{1}{2}S_{1}(t)+\frac{1}{2}S_{2}(t,0)\right)^{\rho}=S_{1}(t)^{\rho},
    \end{split}
\end{align*}
using the fact that  $S_{2}(t,0)=S_{1}(t)$.
We can substitute the results in the Taylor series expansion,
\begin{align*}
    \begin{split}
    f(\Delta)\approx 1+\Delta(S_{1}(t)^{\rho}-S_{1}(t)^{\rho}) = 1
    \end{split}
\end{align*}
The results confirms that the hazard ratio functions are approximately equal to each other when the treatment effect $e^{\Delta}$ is small. 
\section*{\\Appendix C: Taylor series expansion of the true hazard ratio function and the proposed hazard ratio function for the delayed effect effect}

We demonstrate the approximate equality of the two hazard ratio functions for the delayed effect when $\varphi$ is small. We show the results for a general $\gamma$.
Recall, the true hazard ratio and the proposed hazard ratio are of the following form,

\begin{align*}
    \begin{split}
        HR(t,\varphi)&=\frac{L^{\gamma}((\mathcal{L^{\gamma}})^{-1}(\mathcal{L^{\gamma}}(e^{-\lambda_{1}t})+\varphi))}{L^{\gamma}(e^{-\lambda_{1}t})}\\
        HR_{LL}(t,\varphi)&=e^{\beta \left( \frac{\left(1-0.5e^{-\lambda_{1}t}-0.5\mathcal{L}^{-1}(\mathcal{L}(e^{-\lambda_{1}t})+\varphi)\right)^{\gamma}}{\left(1-0.5e^{-\lambda_{1}\tau}-0.5\mathcal{L}^{-1}(\mathcal{L}(e^{-\lambda_{1}\tau})+\varphi)\right)^{\gamma}}\right)}
    \end{split}
\end{align*}
where $L^{\gamma}(x)=\int_{x}^{1} \frac{(1-s)^{\gamma}}{s} ds$ and $\mathcal{L}^{\gamma}(x)=\int_{0.5}^x\frac{1}{sL^{\gamma}(s)} ds$. Recall $S_{2}(t)= (\mathcal{L})^{-1}(\mathcal{L}(e^{-\lambda_{1}t})+ \varphi)$ and $S_{1}(t)=e^{-\lambda_{1}}t$. We again assume that the Lin and León model estimates the maximal treatment effect $\beta$ correctly. For simplicity we write $\beta$ as some function of $\varphi$ and note that the maximal treatment effect occurs at $t=\tau$, so we have $g(\varphi)=\log\left(\frac{L^{\gamma}(S_{2}(\tau,\varphi))}{L^{\gamma}(S_{1}(\tau))}\right)$, 
\begin{align*}
    \begin{split}
        HR_{LL}(t,\varphi)&=e^{g(\varphi) \left( \frac{\left(1-0.5S_{1}(t)-0.5S_{2}(t)\right)^{\gamma}}{\left(1-0.5S_{1}(t)-0.5S_{2}(t)\right)^{\gamma}}\right)}
    \end{split}
\end{align*}

We let $f(\varphi)=\frac{HR(t,\varphi)}{HR_{LL}(t,\varphi)}$ and proceed with the first order Taylor series expansion for the ratio of the hazard ratio functions, i.e. $f(\varphi) \approx f(0)+f'(0)(\varphi - 0)$ for small $\varphi$. We check for $\varphi=0$, $HR(t,0)=\frac{L^{\gamma}(S_{2}(t,0))}{L^{\gamma}(S_{1}(t))}=1$ and $HR_{LL}(t,0)=1$, hence $f(0)=\frac{1}{1}=1$. We now find the first order derivative of $f(\varphi)$ with respect to $\varphi$ and we get the same expression as for the diminishing effect, i.e.
\begin{align*}
    \begin{split}
     f'(0)=\frac{\partial HR(t,0)}{\partial \varphi} -\frac{\partial HR_{LL}(t,0)}{\partial \varphi}   
    \end{split}
\end{align*}
The steps to determine the first order derivative using the quotient rule for the true hazard ratio functions are as follows,
\begin{align*}
    \begin{split}
        HR(t,\varphi)&=\frac{L^{\gamma}(S_{2}(t,\varphi))}{L^{\gamma}(S_{1}(t))}\\
\frac{\partial HR(t,\varphi)}{\partial \varphi} &=\frac{
 L^{'\gamma}(S_{2}(t,\varphi))\frac{\partial S_{2}(t,\varphi)}{\partial \varphi}L^{\gamma}(S_{1}(t))-0}{\left(L^{\gamma}(S_{1}(t))\right)^{2}},
  \end{split}
\end{align*}
Using the inverse function theorem, we have $\frac{\partial S_{2}(t,\varphi)}{\partial \varphi}$=$\frac{1}{\mathcal{L}^{'\gamma}\left((\mathcal{L^{\gamma}})^{-1}(\mathcal{L^{\gamma}}(S_{1}(t))+\varphi)\right)}$. Note that $\mathcal{L}^{'\gamma}(x)=\frac{1}{xL^{\gamma}(x)} $, so we have
\begin{align*}
    \begin{split}
    \frac{\partial S_{2}(t,\varphi)}{\partial \varphi}= (\mathcal{L^{\gamma}})^{-1}\left(\mathcal{L^{\gamma}}(S_{1}(t))+\varphi\right) L^{\gamma}\left((\mathcal{L^{\gamma}})^{-1}(\mathcal{L^{\gamma}}(S_{1}(t))+\varphi)\right)
    \end{split}
\end{align*}
We now evaluate $\frac{\partial HR(t,\varphi)}{\partial \varphi}$ at $\varphi=0$,
\begin{align*}
    \begin{split}
       \frac{\partial HR(t,0)}{\partial \varphi}&=\frac{L^{'\gamma}(S_{2}(t,0))S_{1}(t)L^{\gamma}(S_{1}(t)) L^{\gamma}(S_{1}(t))}{(L^{\gamma}(S_{1}(t)))^{2}} \\
       &=L^{'\gamma}(S_{1}(t))S_{1}(t)
    \end{split}
\end{align*}
Note that $L^{'\gamma}(x)=\frac{(1-x)^{\gamma}}{x}$. We then have, 
\begin{align*}
    \begin{split}
      \frac{\partial HR(t,0)}{\partial \varphi}&=\frac{(1-S_{1}(t))^{\gamma}}{S_{1}(t)}S_{1}(t)=(1-S_{1}(t))^{\gamma}
    \end{split}
\end{align*}

Now we determine the first order derivative for the proposed hazard ratio function using the product and chain rule,
\begin{align*}
    \begin{split}
     HR_{LL}(t,\varphi)&=e^{g(\varphi) \left( \frac{\left(1-0.5S_{1}(t)-0.5S_{2}(t,\varphi)\right)^{\gamma}}{\left(1-0.5S_{1}(\tau)-0.5S_{2}(\tau,\varphi)\right)^{\gamma}}\right)} \\
    \frac{\partial HR_{LL}(t,\varphi)}{\partial \varphi}&=HR_{LL}(t,\varphi) \times \Bigg\{ g'(\varphi) \frac{\left(1-0.5S_{1}(t)-0.5S_{2}(t,\varphi)\right)^{\gamma}}{\left(1-0.5S_{1}(\tau)-0.5S_{2}(\tau,\varphi)\right)^{\gamma}} 
     + \\&g(\varphi)
    \left(\frac{\left(1-0.5S_{1}(t)-0.5S_{2}(t,\varphi)\right)^{\gamma}}{\left(1-0.5S_{1}(\tau)-0.5S_{2}(\tau,\varphi)\right)^{\gamma}}\right)' \Bigg\} \\
    \frac{\partial HR_{LL}(t,0)}{\partial \varphi}&=HR_{LL}(t,0)\times \Bigg\{ g'(0)\frac{\left(1-S_{1}(t)\right)^{\gamma}}{\left(1-S_{1}(\tau)\right)^{\gamma}} 
     + \\&g(0)
    \left(\frac{\left(1-0.5S_{1}(t)-0.5S_{2}(t,\varphi)\right)^{\gamma}}{\left(1-0.5S_{1}(\tau)-0.5S_{2}(\tau,\varphi)\right)^{\gamma}} \right)' \Bigg\}
        \end{split}
    \end{align*}
Note that $g(0)=\log\left(\frac{L^{\gamma}(S_{2}(\tau,0))}{L^{\gamma}(S_{1}(\tau))}\right)=\log(1)=0$. We now determine $g'(0)$,
\begin{align*}
    \begin{split}
   g(\varphi)&= \log\left(L^{\gamma}(S_{2}(\tau,\varphi))\right)-\log\left({L^{\gamma}(S_{1}(\tau))}\right)\\
   g'(\varphi)&= \frac{1}{L^{\gamma}(S_{2}(\tau,\varphi))} \times L^{'\gamma}(S_{2}(\tau,\varphi)) \times (\mathcal{L^{\gamma}})^{-1}\left(\mathcal{L^{\gamma}}(S_{1}(\tau))+\varphi\right) L^{\gamma}\left((\mathcal{L^{\gamma}})^{-1}(\mathcal{L^{\gamma}}(S_{1}(\tau))+\varphi)\right) \\
   g'(0)&=\frac{1}{L^{\gamma}(S_{1}(\tau))} \times L^{'\gamma}(S_{2}(\tau,0)) \times S_{1}(\tau)L^{\gamma}(S_{1}(\tau)) \\
   &=\frac{\left(1-S_{1}(\tau)\right)^{\gamma}}{S_{1}(\tau)}S_{1}(\tau)=\left(1-S_{1}(\tau)\right)^{\gamma}\\
      \end{split}
\end{align*}
Substituting the results in $\frac{\partial HR_{LL}(t,0)}{\partial \varphi}$ gives,
 \begin{align*}
    \begin{split}
    \frac{\partial HR_{LL}(t,0)}{\partial \varphi}&=e^{0}\times\left\{ \left(1-S_{1}(\tau)\right)^{\gamma} \frac{\left(1-S_{1}(t)\right)^{\gamma}}{\left(1-S_{1}(\tau)\right)^{\gamma}} + 0\right\} \\
    &=\left(1-S_{1}(t)\right)^{\gamma}
    \end{split}
\end{align*}
We can substitute the results in the Taylor series expansion, 
\begin{align*}
    \begin{split}
        f(\varphi)\approx 1+\varphi\left\{\left(1-S_{1}(t)\right)^{\gamma}-\left(1-S_{1}(t)\right)^{\gamma}\right\}=1
    \end{split}
\end{align*}
This result confirms that the hazard ratio functions are approximately equal to each other when the treatment effect is small in the case of delayed effect. 
\newpage

\bibliographystyle{unsrt}
\bibliography{references.bib}

\end{document}